\documentclass{emulateapj}

\usepackage{natbib}
\usepackage{amsmath}
\usepackage{color}

\newcommand{\Ni}{$^{56}$Ni}
\newcommand{\Co}{$^{56}$Co}

\shorttitle{Dip after Early Emission of SLSNe}
\shortauthors{Moriya \& Maeda}

\begin{document}

\title{
A DIP AFTER THE EARLY EMISSION OF SUPER-LUMINOUS SUPERNOVAE: \\
A SIGNATURE OF SHOCK BREAKOUT WITHIN DENSE CIRCUMSTELLAR MEDIA
}

\def\ipmu{1}
\def\ut{2}
\def\resceu{3}

\author{
{Takashi J. Moriya}\altaffilmark{\ipmu,\ut,\resceu} and
{Keiichi Maeda}\altaffilmark{\ipmu}}

\altaffiltext{\ipmu}{
Kavli Institute for the Physics and Mathematics of the Universe,
Todai Institutes for Advanced Study,
University of Tokyo, Kashiwanoha 5-1-5, Kashiwa, Chiba 277-8583, Japan;
takashi.moriya@ipmu.jp
}
\altaffiltext{\ut}{
Department of Astronomy, Graduate School of Science, University of Tokyo,
Hongo 7-3-1, Bunkyo, Tokyo 113-0033, Japan
}
\altaffiltext{\resceu}{
Research Center for the Early Universe,
Graduate School of Science, University of Tokyo,
Hongo 7-3-1, Bunkyo, Tokyo 113-0033, Japan
}

\begin{abstract}
The origin of super-luminous supernovae (SLSNe), especially the source
 of their huge luminosities, has not been clarified yet. While a strong
 interaction between SN ejecta and dense circumstellar media (CSM)
 is a leading scenario, alternative models have been proposed. In
 this {\em Letter}, we suggest new diagnostics to discriminate the
 strong SN-CSM interaction scenario from the others: a decline
 in the luminosity ('dip') before the main peak of the light curve.
This dip is an unavoidable consequence of having a dense CSM
within which the shock breakout occurs.
 If a dense CSM shell is located far at large radii from the progenitor
 inside, it takes time for the SN ejecta to reach it
and the early light curve can be powered by
the SN ejecta before the collision.
Once the SN
 ejecta collides with the dense CSM, the electron density and
 thus the Thomson scattering opacity suddenly increase.
 Photons become unable to go out of the shock even if there is a
 source of emission inside, which results in the dip in the
 light curve. This dip is a solid prediction from the strong interaction
 scenario irrespective of a power source for the early emission. Eventually
 the forward shock breaks out from within the dense
 CSM, and the luminosity increases by the continuous strong SN-CSM
 interaction, resulting in an SLSN. The possible dip observed in the
 hydrogen-poor SLSN, 2006oz, could be the first
 example of this signature and give support to the SN-CSM interaction
 as the power source of SLSN 2006oz.
\end{abstract}


\section{Introduction}
The origin of the huge luminosities of super-luminous supernovae (SLSNe), 
categorized by maximum luminosities exceeding $\sim 10^{44}~\mathrm{erg~s^{-1}}$, 
is one of the biggest mysteries in the study of stellar explosions. 
SLSNe are roughly divided into two groups based on the existence or
non-existence of hydrogen lines in their spectra (H-rich and H-poor
SLSNe). The origin of H-rich SLSNe is most likely a strong interaction
between the SN ejecta and a dense circumstellar medium (CSM)
\citep[e.g.,][]{woosley2007,chevalier2011,moriya2012,svirski2012,moriya2012b}
because many such events show narrow H emission lines
(i.e., they are Type IIn SNe) which
indicates the existence of a dense CSM \citep[e.g.,][]{smith2010}. On the other hand, the origin of H-poor SLSNe is not yet well-understood.
Some of them show light curves (LCs) whose decline rates after the peak
are consistent with the \Co\ decay and they are likely powered by a
large amount of \Ni\
\citep[e.g.,][]{gal-yam2009,young2010,moriya2010}. However, this is not
the case for the majority of H-poor SLSNe. Their LCs generally decline
much faster than the \Co\ decay
\citep{quimby2007,quimby2011,barbary2009,pastorello2010,chomiuk2011}. For
these H-poor SLSNe, several mechanisms to power the LCs have been
suggested, including an interaction of SN ejecta with C+O-rich dense CSM
\citep[e.g.,][]{blinnikov2010}, a spin-down of a highly-magnetized
neutron star \citep[e.g.,][]{kasen2010,woosley2010,maeda2007}, or a
quark nova model \citep[e.g.,][]{ouyed2012}.
However, there has been no clear observational evidence to
distinguish the actual heating mechanism of H-poor SLSNe.

Recent observations of an H-poor SLSN 2006oz revealed the existence of
the early emission before the main part of the LC \citep{oz}. The early emission
was observed to continue about 10 days at the bolometric
luminosity of $\sim 10^{43}$ erg s$^{-1}$, followed by a possible
decline in the luminosity for a few days (a 'dip'). Then the luminosity
increased to at least $\sim 10^{44}$ erg s$^{-1}$ in the timescale of
$\simeq 30$ days (`main LC'). It is unlikely that emission from SN
2006oz was powered by the \Ni\ decay, since most of the ejecta $(\simeq
10~M_\odot)$ would have to be \Ni\ to simultaneously explain the rising time
and the peak luminosity of SN 2006oz with \Ni\ heating \citep{oz}.
The origins of the
early emission and the possible dip in the LC have not yet been clarified \citep{oz}. 

In this {\em Letter}, we explore a consequence of the SN-dense CSM
interaction scenario for SLSNe.
We show that this scenario predicts 
that the luminosity of SLSNe should decline for a while
before the strong interaction that powers the huge luminosity begins.
The dip is an inevitable consequence of the shock breakout within the
dense CSM (\citealt{chevalier2011,moriya2012,svirski2012}, see also
\citealt{ofek2010,balberg2011}). The possible dip observed in SN 2006oz
could be the first example of this and it indicates that the SN-CSM interaction
is the power source of the H-poor SLSN 2006oz.

\section{Dense Circumstellar Medium around SN 2006oz}\label{CSM}
We explore a consequence of the SN-dense CSM
interaction scenario to power the emission from SLSNe. Although our
arguments apply to any SLSNe powered by interaction, we focus on 
SN 2006oz to provide our basic idea. This SLSN is the
best example so far for which the early phase behavior has been well
observed. First, we estimate physical properties of the dense CSM around the progenitor of SN 2006oz, under the assumption that the main LC was powered by the interaction
between SN ejecta and dense CSM. Then, with these constraints, we discuss
what is expected to take place in the proposed system before the main LC. 

Figure \ref{fig1}a presents the progenitor system required in the strong
interaction scenario. A dense CSM shell exists between $R_i$ and $R_o$.
Once the SN ejecta reaches $R_i$, the strong
interaction takes place until the ejecta reaches $R_o$, and
this interaction powers the main LC.
An early emission is created in the phase before the ejecta reaches
to $R_i$. Our main arguments below do not depend on the nature of a
power source for the early emission, and thus we proceed without specifying
it (see Section \ref{conc} for possible origins for the early emission). The CSM
should be dense enough to explain the peak luminosity of SN 2006oz
by the interaction scenario, and the shock breakout is
expected to take place within the CSM at the beginning of the strong interaction. The radius where the shock breakout occurs is expressed as $xR_o$
(where $R_i/R_o<x<1$). As we focus on H-poor SLSNe, we assume that the
CSM is mainly composed of C and O, and the progenitor star is a
Wolf-Rayet (WR) star. In the following, we assume that the dense CSM is
uniformly distributed with a constant density. This is just for the sake
of simplicity, and the main result is not sensitive to this
assumption. Under these assumptions, we estimate properties of the dense
CSM by comparing the shock breakout prediction and the observed features of SN 2006oz. 

The blackbody radius obtained from the spectrum near the main LC peak of
SN 2006oz is about $2.5\times 10^{15}$ cm \citep{oz}. Since the last
scattering surface of the CSM at the main LC peak is expected to be at
the outermost region of the dense CSM shell when the density is constant
in the dense CSM \citep{moriya2012}, we can estimate that $R_o\simeq 2.5\times 10^{15}$ cm. On the other hand, the 
blackbody radius at the beginning of the main LC rising phase (i.e., at the beginning of the strong interaction just after the shock breakout within the CSM) is $\simeq 10^{15}$ cm \citep{oz}. Thus, we can estimate that $xR_o\simeq 10^{15}$ cm. 

By assuming that the rising time of the main LC of SN 2006oz ($\simeq30$ days)
corresponds to the diffusion time $t_d$ of the dense CSM,
the electron density $n_e$ in the dense CSM can be estimated from the following equation: 
\begin{equation}
t_d\simeq\frac{\tau_T (R_o-xR_o)}{c},\label{eq1}
\end{equation}
where $\tau_T=\sigma_T n_e (R_o-xR_o)$ is the Thomson scattering optical depth
within the dense CSM, $c$ is the speed of light, and $\sigma_T$ is the Thomson
scattering cross section. From Equation (\ref{eq1}), 
\begin{equation}
n_e\simeq\frac{ct_d}{\sigma_T(R_o-xR_o)^2}\simeq 5\times 10^{10}~\mathrm{cm^{-3}}.
\label{ne}
\end{equation}
The last value of Equation (\ref{ne}) is estimated by adopting the
parameters for SN 2006oz, i.e.,
$t_d=30$ days, $xR_o=10^{15}$ cm, and $R_o=2.5\times 10^{15}$ cm.
$\tau_T=52$ in this case and it is plausible that
the shock breakout occurs
in the CSM with the typical forward shock velocity
$v_s \simeq 10,000~\mathrm{km~s^{-1}}$ $(c/v_s\simeq 30)$.

If the dense CSM is composed of 50 \% C and 50 \% O and both C and O are singly ionized in the entire CSM, the CSM density corresponding to $n_e\simeq 5\times 10^{10}~\mathrm{cm^{-3}}$
is $\rho_{\rm CSM} \simeq 10^{-12}~\mathrm{g~cm^{-3}}$. Then, the
required CSM mass is $\simeq 35~M_\odot$. If we further assume that the
outflowing velocity of the dense CSM was $100~\mathrm{km~s^{-1}}$, the
$35~M_\odot$ of C+O-rich materials must have been lost from the
progenitor within 8 years before the explosion at a rate of $\simeq
7~M_\odot~\mathrm{yr^{-1}}$. Mechanisms by which WR stars experience
such a huge mass loss just before the explosion have not yet been
clarified, although there are some suggestions
\citep[e.g.,][]{quataert2012}. Alternatively, the dense CSM does not
necessarily need to come from the huge mass ejection
from the progenitor. Within a dense cluster, collisions of WR stars may
leave dense C+O-rich envelopes that would persist until the time of the
explosion. This is an alternative way to have a dense C+O-rich CSM
around an SN \citep[see also, e.g.,][]{portegies2007,pan2011,chevalier2012}.

\begin{table}
\centering
\caption{Model Parameters of SN 2006oz}
\label{table1}
\begin{tabular}{cccccc}
\hline
$R_i$ & $xR_o$ & $R_o$ & $n_e$ & CSM Density &CSM Mass \\
cm & cm & cm & $\mathrm{cm^{-3}}$ & $\mathrm{g~cm^{-3}}$ &$M_\odot$ \\
\hline
$10^{15}$ &$10^{15}$& $2.5\times 10^{15}$& $5\times 10^{10}$&$10^{-12}$ &$35$ \\
\hline
\end{tabular}
\end{table}

\section{A dip as a signature of the SN-CSM interaction}\label{sec:dip}
Based on the properties of the dense CSM required by the SN-CSM
interaction scenario to power the main LC of SLSN 2006oz discussed
above,
we now investigate a consequence of this scenario in the early phase
before the main LC. We
suggest that there must be a brief phase of decreased luminosity lasting for
a few days before the strong interaction energizes the main LC.
This argument is independent of any assumptions
regarding the nature of the early emission which will be discussed in Section
\ref{conc}. The only requirement is that there is a detectable, i.e.,
sufficiently luminous, early emission phase. The dip phase should then
appear as the fading phase between the early emission and the main LC. 

\begin{figure*}[b]
\begin{center}
\includegraphics[width=0.45\columnwidth]{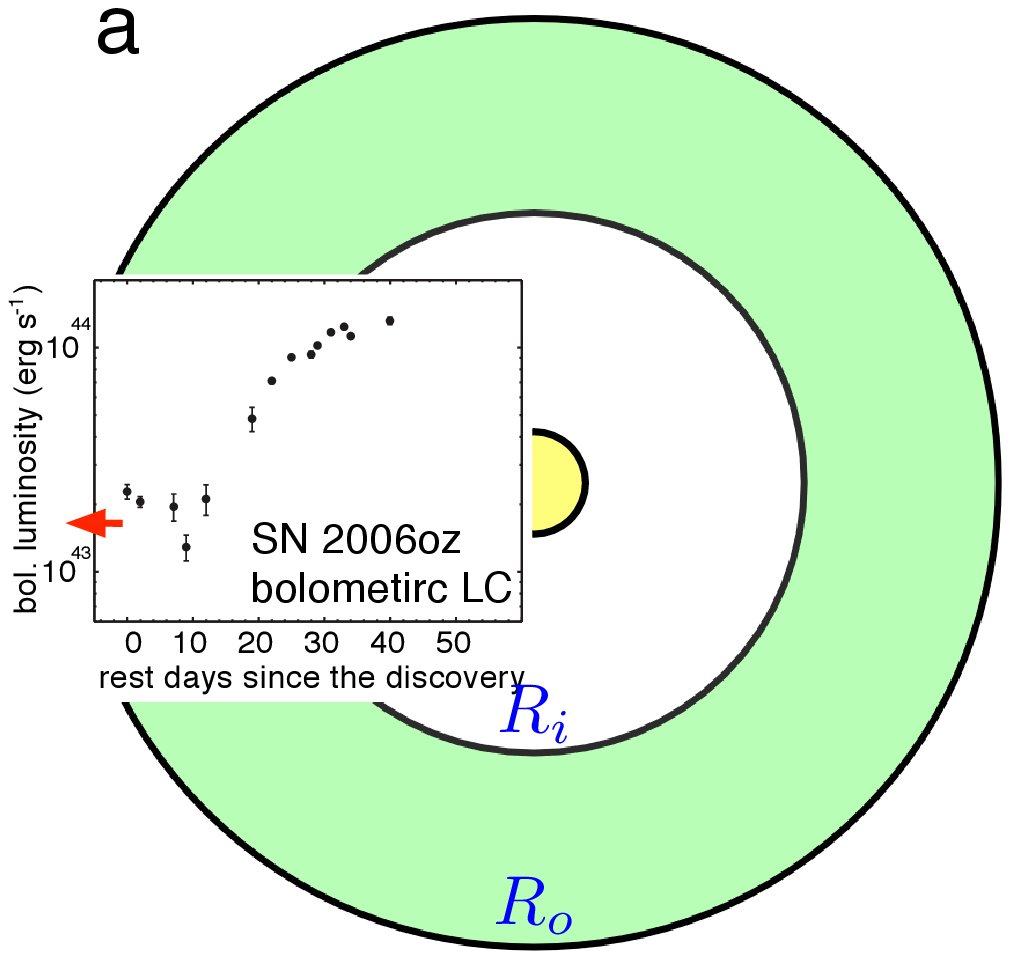}
\includegraphics[width=0.45\columnwidth]{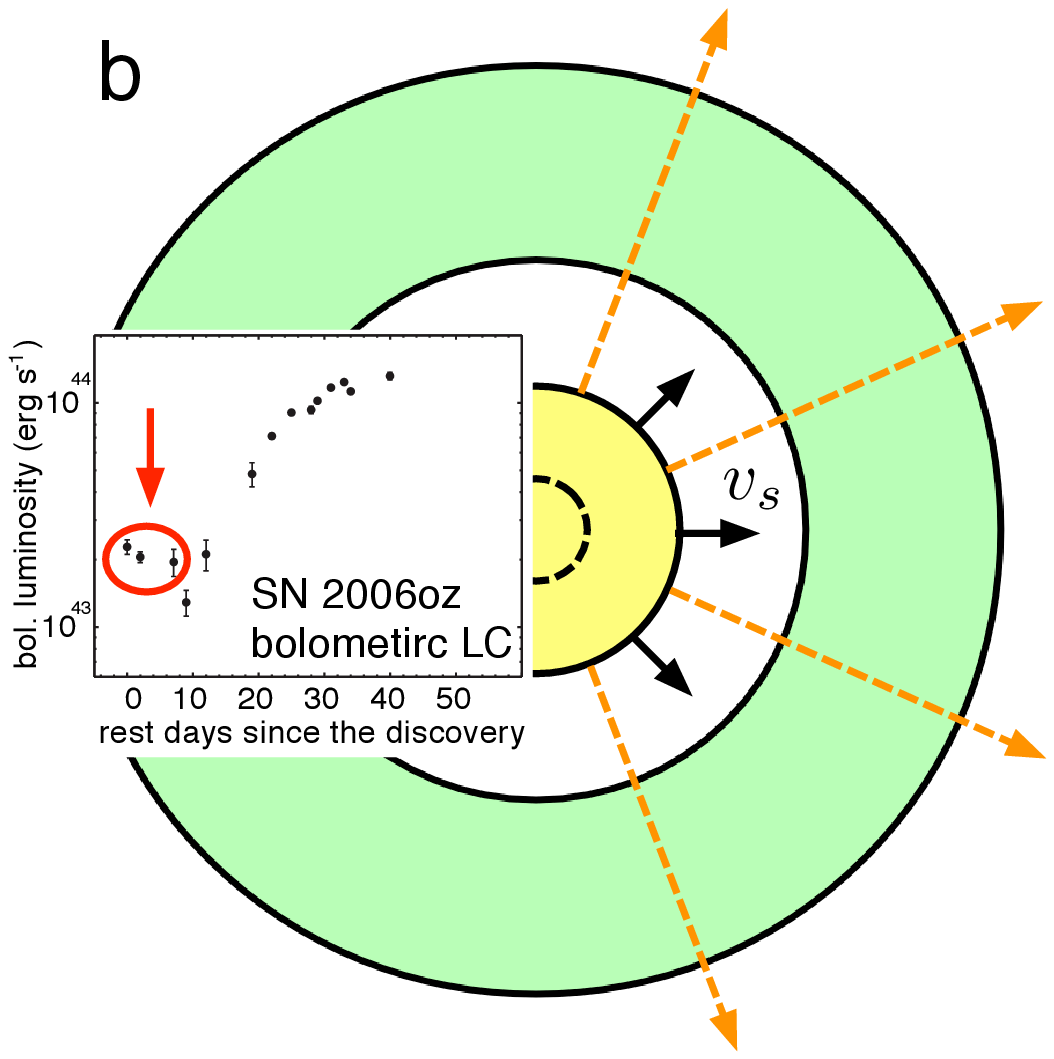}\\
\includegraphics[width=0.45\columnwidth]{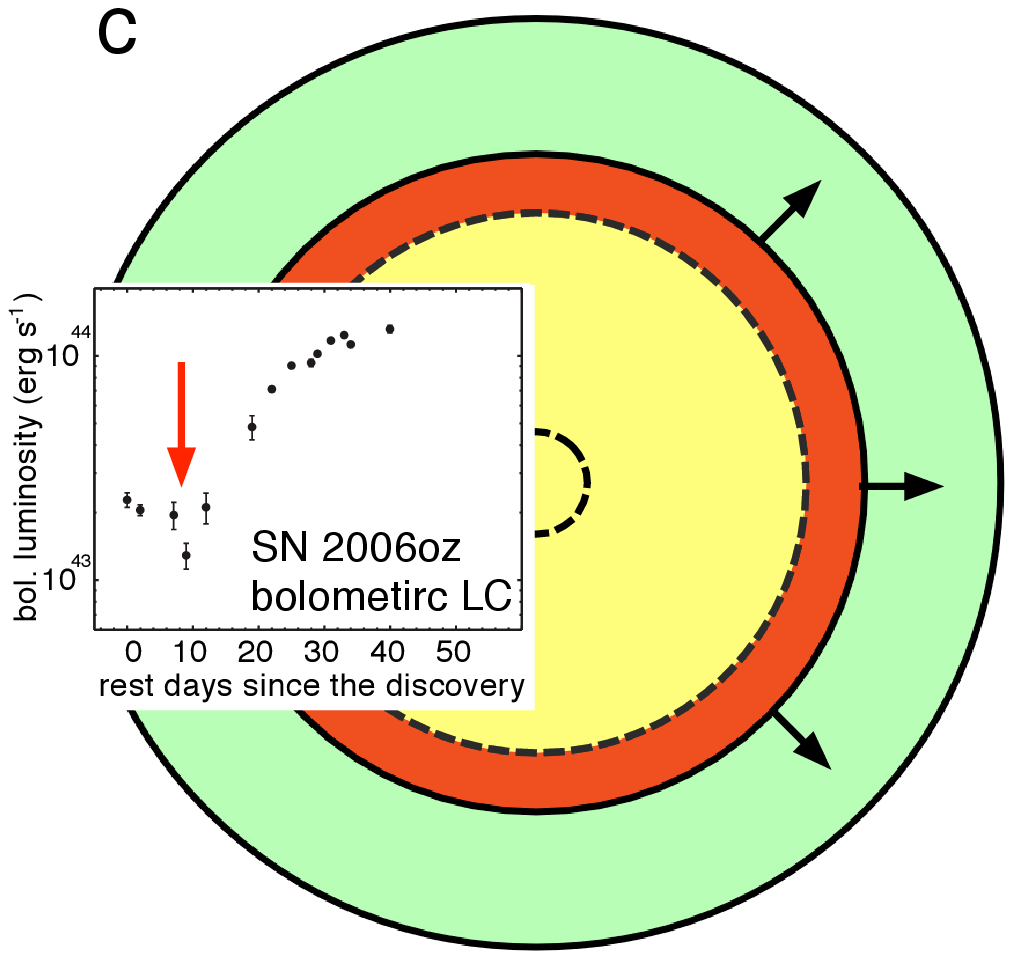}
\includegraphics[width=0.45\columnwidth]{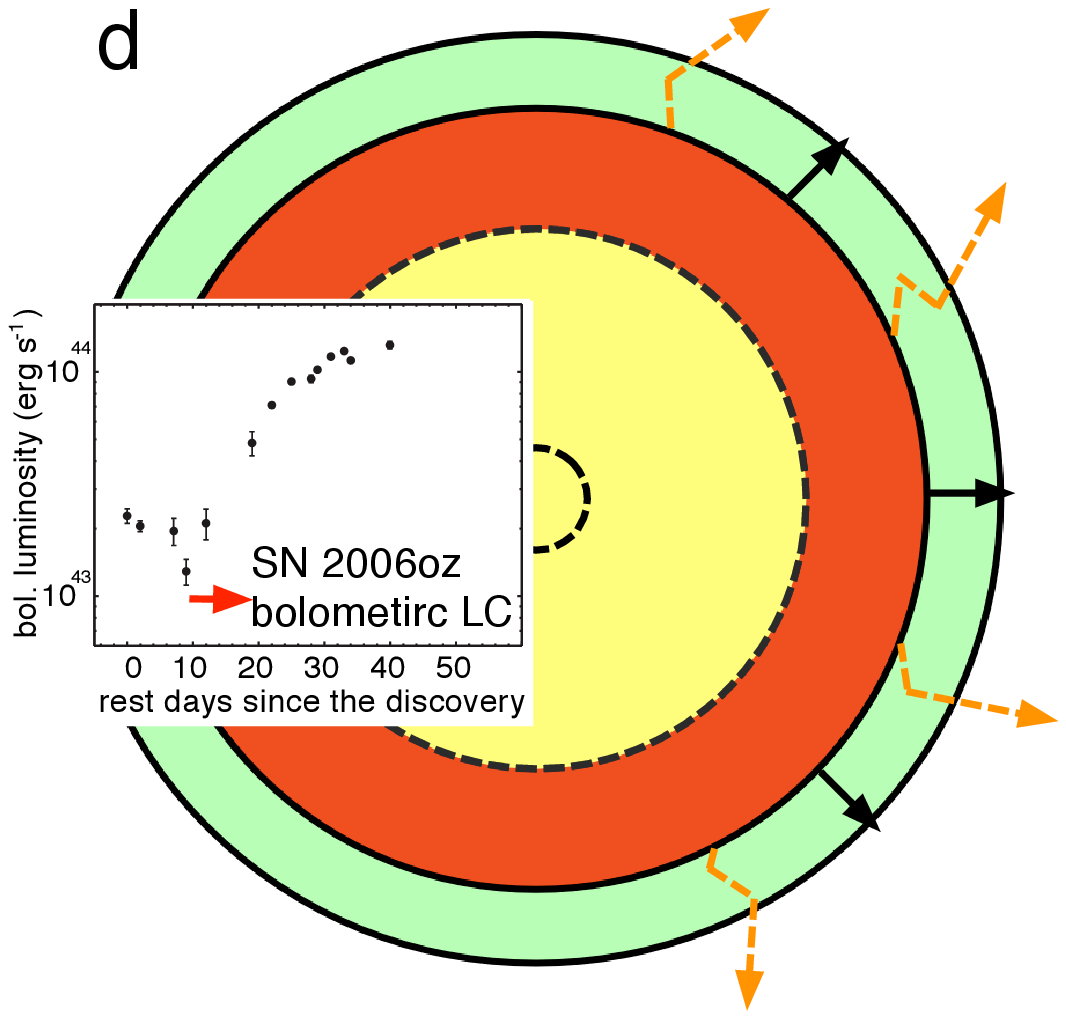}
\caption{
The proposed origin of the dip between the early emission and the main LC.
The dense CSM, within which the shock breakout occurs, extends
from $R_i$ to $R_o$. A progenitor star explodes in this configuration (a).
Before the forward shock 
reaches $R_i$ (b), the dense CSM is transparent to optical photons
 from the SN ejecta and they can be observed as an early emission. When the
 forward shock reaches $R_i$ (c), the opacity suddenly increases
 as ionization in the dense CSM is enhanced. Thus, even if the
 energy source of the early emission is still active, the dense CSM blocks
 the light, causing a dip in the LC. Then, after the shock breakout
 within the dense CSM (d), photons can escape from the shock. The SN is now powered by the strong interaction reaching the large luminosity in the main peak. The LC shown in each inset is the bolometric LC of SN 2006oz obtained by \citet{oz}.
}\label{fig1}
\end{center}
\end{figure*}

Figure \ref{fig1} summarizes our model for the dip after the early
emission.  Before the explosion, most of C and O in the dense CSM is not
ionized, and thus the CSM is transparent. This is because of the high
CSM density which results in the high recombination rate ($\simeq 10^{-12}~\mathrm{g~cm^{-3}}$ estimated for SN 2006oz). The emission rate of the ionizing photons from a typical WR star ($10^6L_\odot$ and $10^5$ K)  is $\sim 10^{49}~\mathrm{s^{-1}}$. With the recombination coefficient $\sim 10^{-13}~\mathrm{cm^{3}~s^{-1}}$, the number of ionizing photons is too small to keep the dense CSM ionized. 

Then, the central star explodes as an SN. Before the ejecta reaches $R_i$ (Figure \ref{fig1}b), 
the SN ejecta expands within the rarefied region below $R_i$.
We attribute the early emission in the LC to the light from the SN ejecta in
this phase before the strong collision.
The duration of the early emission in SN 2006oz before the main LC is about
10 days. Regardless of the nature of the early emission, the duration can be interpreted as the time required for the SN ejecta to reach the dense CSM (i.e., at $R_i$) in our scenario. With $v_s\simeq10,000~\mathrm{km~s^{-1}}$, it reaches $\simeq 10^{15}$
cm in about 10 days and we can estimate that $R_i\simeq xR_o$. This is
consistent with the estimated blackbody radius in this phase (Section \ref{CSM}). 

Regardless of the mechanism to power the early emission, if the majority of
photons emitted from the SN ejecta is in the optical or near-ultraviolet,
most of C and O in the dense CSM will still not be ionized during the
early emission phase. For instance, the blackbody radius and temperature of
SN 2006oz during the early emission phase
are $\simeq 10^{15}$ cm and $15,000$ K,
respectively, and the emission rate of the ionizing photons ($\sim
10^{54}~\mathrm{s^{-1}}$) is too small to keep most of the dense CSM
ionized. Only the innermost thin layer of the dense CSM (up to $\simeq
1.2\times 10^{15}$ cm) can be ionized in this case.
The optical depth to the Thomson scattering in this ionized region becomes
$\simeq 7$. However, the ionizing region is confined in the thin layer,
and the diffusion time scale within it is estimated to be less than a
day. Thus, the photosphere can be located in this thin layer of
ionized material, but the effect on the LC evolution is expected to be
small. Some recombination lines may be found in spectra at this
phase. Nonetheless, most of the dense CSM is still transparent to
optical photons, and thus we can observe the early emission. 

After about 10 days after the explosion, the SN ejecta starts to collide with the dense CSM.
Because of the strong interaction, X-rays and ultraviolet photons are
now efficiently produced at the forward shock\footnote{
As X-rays can reach innter orbit electrons, they may have
difficulties to propagate outward.},
and the electron density
in the dense CSM suddenly increases. The CSM gets ionized and the
Thomson scattering makes the dense CSM opaque to any photons.
Then, the diffusion velocity of photons can be less than the velocity
of the shock wave, and photons cannot go out of the shock
until the shock breakout takes place
at $xR_o$ (Figure \ref{fig1}c). During this optically thick phase before
the shock breakout, the luminosity decreases. This sudden decline in the luminosity is a naturally-expected observable signature of the strong SN-CSM interaction scenario. 

We suggest that the possible dip observed in SN 2006oz could be the
first observed example of this signature. The duration of the dip in the
LC of SN 2006oz was short. The luminosity at the single observed epoch after the early emission showed the decline, and the luminosity was back to the previous level by the next epoch (\citealt{oz}, Figure \ref{fig1}). Therefore, the duration of the dip was at most 2 days. From this, we can place a constraint on the shock breakout: $xR_o-R_i$ should be less than $2\times 10^{14}$ cm, or $xR_o$ should be less than $1.2\times 10^{15}$ cm, if $v_s=10,000~\mathrm{km~s^{-1}}$. We note that the duration of the dip can be very short, and thus high cadence observations are important to capture this signature. 

After the shock breakout at $xR_o$, photons are able to escape out of
the interaction region. Then, the SN luminosity is powered by the SN-CSM
interaction, and the SN becomes super-luminous from the ongoing
strong interaction (Figure \ref{fig1}d, Section \ref{CSM}).

\section{Discussion and Conclusions}\label{conc}
We suggest a new way to distinguish proposed power sources
of SLSNe. Among scenarios proposed so far, the strong interaction
scenario, which requires the existence of dense CSM, is distinguishable by
the early phase LC before the rising part to the peak luminosity. The
scenario predicts that a brief dip phase should appear
before the main LC if there is an early emission which is bright enough to be
observed, as was the case for the H-poor SLSN 2006oz.
This argument is irrespective of detailed nature and origin of the
early emission itself. The existence of the dip reflects the change in the
ionization condition in the dense CSM following the SN-CSM interaction
which results in the shock breakout within the CSM. The possible dip observed in H-poor SLSN 2006oz indicates that the main power source of the huge luminosity for this SLSN is the strong interaction between the SN ejecta and the dense C+O-rich CSM whose mass is estimated as $\simeq 35~M_\odot$.
Other proposed power sources like magnetars may also happen to show a
dip for a specific combination of model parameters and a dip may
appear in some SLSNe but not in all SLSNe in these scenarios.
On the other hand, a dip should always appear when the shock breakout occurs.
Thus, more H-poor SLSN samples in the early phase are required to
see whether a dip is a common feature of H-poor SLSNe and it is
actually due to the shock breakout. We strongly encourage future
observations in this direction.

The early emission of SN 2006oz itself is bright $(\sim
10^{43}~\mathrm{erg~s^{-1}})$, with the total radiation energy of $\sim
10^{49}$ erg within $\simeq$ 10 days. There are a few possible mechanisms
to power the early emission. \Ni\ produced in the SN inside is one
possibility. The color of the early emission obtained by \citet{oz} is
similar to that of Type Ia SNe near the LC peak
\citep[e.g.,][]{wang2009}. The required \Ni\ mass to explain the
early emission luminosity is $\sim 1~M_\odot$. However, a difficulty in this
model is that the rising time of the early emission is constrained to be at
most 5 days \citep{oz}, which is too short for the \Ni\ heating
scenario. Another possibility is the interaction between the SN ejecta
and CSM. It is possible that CSM which is less dense than the dense CSM
above $R_i$ exists below $R_i$. If there is additional CSM of $\sim
0.1~M_\odot$ below $R_i$, this is enough to create the luminosity of
$\sim 10^{43}~\mathrm{erg~s^{-1}}$ through the SN-CSM interaction
\citep{moriya2011}. This small amount of CSM would not change the
overall picture we suggest, since the total amount of the radiation
energy emitted as the early emission $(\sim 10^{49}~\mathrm{erg})$ is much
smaller than the total available kinetic energy ($\sim 10^{51}$ erg or
even more) by the SN explosion and does not affect the dynamics of the
shock wave so much. 

The existence of $\sim 10~M_\odot$ C+O-rich CSM
around a WR star which is lost just before its explosion
clearly challenges the current understanding of stellar mass loss and
stellar evolution. This drastic mass loss could influence the final
progenitor mass at the time of its explosion and its fate.
For example, stars which are currently considered to end up with
a black hole due to fallback may actually become a neutron star
because of the extra mass loss which reduces the mass of
the accreting envelope material at the time of the core collapse.
We still do not have a large number of observations
to confirm that a WR star can actually have such mass loss and
the dip is a common feature of H-poor SLSNe.
Future observations of H-poor SLSNe especially in the early phases
are essential for understanding the origin of H-poor SLSNe and the final fates of
WR stars.

\begin{acknowledgments} 
We thank the anonymous referee for the valuable comments.
We also thank Robert Quimby and Giorgos Leloudas
for their comments on the manuscript.
T.J.M. is supported by the Japan Society for the Promotion of Science
 Research Fellowship for Young Scientists ($23\cdot5929$).
K.M. acknowledges financial support by Grant-in-Aid for Scientific Research
for Young Scientists (23740141).
This research is also supported by World Premier International Research
 Center Initiative, MEXT, Japan.
\end{acknowledgments}

\bibliographystyle{apj}

\end{document}